\documentclass{emulateapj}
\usepackage{graphicx}
\usepackage{amssymb}
\usepackage{amsmath}

\newcommand\be{\begin{equation}}
\newcommand\ee{\end{equation}}

\shorttitle{}
\shortauthors{}

\begin{document}

\title{Long-Lived Dust Asymmetries at Dead Zone Edges in Protoplanetary Disks}

\author{Ryan Miranda{$^{1,2}$}, Hui Li{$^2$}, Shengtai Li{$^2$}, and Sheng Jin{$^{3,2}$}}
\affil{{$^1$} Cornell Center for Astrophysics and Planetary Science, Department of Astronomy, Cornell University, Ithaca, NY 14853, USA \\
{$^2$} Theoretical Division, Los Alamos National Laboratory, Los Alamos, NM 87545, USA \\
{$^3$} Key Laboratory of Planetary Sciences, Purple Mountain Observatory, Chinese Academy of Sciences, Nanjing 210008, China}
\email{rjm456@cornell.edu}

\begin{abstract}
A number of transition disks exhibit significant azimuthal asymmetries in thermal dust emission. One possible origin for these asymmetries is dust trapping in vortices formed at the edges of dead zones. We carry out high-resolution, two-dimensional hydrodynamic simulations of this scenario, including the effects of dust feedback. We find that, although feedback weakens the vortices and slows down the process of dust accumulation, the dust distribution in the disk can nonetheless remain asymmetric for many thousands of orbits. We show that even after $10^4$ orbits, or $2.5$ Myr when scaled to the parameters of Oph IRS 48 (a significant fraction of its age), the dust is not dispersed into an axisymmetric ring, in contrast to the case of a vortex formed by a planet. This is because accumulation of mass at the dead zone edge constantly replenishes the vortex, preventing it from being fully destroyed. We produce synthetic dust emission images using our simulation results. We find that multiple small clumps of dust may be distributed azimuthally. These clumps, if not resolved from one another, appear as a single large feature. A defining characteristic of a disk with a dead zone edge is that an asymmetric feature is accompanied by a ring of dust located about twice as far from the central star.
\end{abstract}

\keywords{protoplanetary disks -- hydrodynamics -- submillimeter: planetary systems}

\section{Introduction}
A number of transitions disks (protoplanetary disks with central dust cavities; see Espaillat et al.~2014 for a review) exhibit significant asymmetries in mm/sub-mm dust emission (van der Marel et al.~2013; Casassus et al.~2013; Isella et al.~2013; P{\'e}rez et al.~2014; van der Marel et al.~2016). It is commonly suggested that these asymmetries are the result of dust trapping in large-scale vortices (e.g., Reg{\'a}ly et al.~2012; Lyra \& Lin 2013; Zhu \& Stone 2014). These can arise at axisymmetric ``bumps'' in the disk, as a result of the Rossby Wave Instability (RWI; Lovelace et al.~1999; Li et al.~2000, 2001; M{\'e}heut et al.~2012a). Note that an alternative origin for asymmetries, resulting from the presence of a central binary, was recently suggested by Ragusa et al.~(2017).

One route to vortex formation is the opening of a gap by a planet embedded in the disk. The outer edge (and sometimes inner edge) of the gap can be RWI-unstable, resulting in the formation of a vortex (Li et al.~2005). The survival of the vortex is inhibited by both viscosity, and by feedback drag exerted on the gas by the accumulated dust (Fu et al.~2014a, 2014b), although continuous accretion can help to sustain them. Unless both the disk viscosity and dust-to-gas ratio are very small, a vortex formed in this fashion is unlikely to survive for more than a few thousand orbits, after which the vortex is dissipated, and the dust contained within it is dispersed into a ring (Fu et al.~2014b; Surville et al.~2016). For the typical orbital periods of observed disk asymmetries, this vortex lifetime corresponds to a small fraction of the disk lifetime, which may be several times $10^4$ orbits. This implies that it is unlikely to be observed, which is problematic if disk asymmetries are common.

There is an alternative channel for producing dust-trapping vortices via the RWI that does not require embedded planets. Protoplanetary disks are expected to be inefficient at transporting angular momentum via turbulence in their inner regions, due to suppression of the magnetorotational instability by non-ideal MHD effects (Gammie 1996; Bai \& Stone 2013; Bai 2014; Lesur et al.~2014; Simon et al.~2015; Bai 2016), and a lack of non-thermal ionization sources (e.g., Cleeves et al.~2013). This region, known as the dead zone (DZ), may extend to a significant fraction of $100 \mathrm{AU}$ from the central star. Beyond the DZ, the disk is turbulent, resulting in angular momentum transport and mass accretion. The edge of the DZ is therefore characterized by a sharp increase in turbulent viscosity (for the case of an Ohmic DZ, the gradient of effective viscosity at the DZ edge is sharp, even when that of the underlying resistivity is not; Lyra et al.~2015). The presence of a viscosity transition leads to accumulation of mass, creating an RWI-unstable bump at which a vortex may be formed (Varni{\`e}re \& Tagger 2006; Lyra et al.~2009; Reg{\'a}ly et al.~2012; Lyra \& Mac Low 2012; Miranda et al.~2016). Note that the DZ also has an inner edge, but its proximity to the central star ($\sim 0.1 \mathrm{AU}$; Gammie 1996) makes it irrelevant to the dynamics of the outer disk ($\sim 50 \mathrm{AU}$) where asymmetric features are observed, and so we are only concerned with the outer edge of the DZ in this study.

In this paper, we carry out high-resolution 2D hydrodynamic simulations of vortex formation at a DZ edge, including a full treatment of dust dynamics, with feedback. We evolve the disk for $10^4$ orbits, representing a substantial fraction of the lifetime of a protoplanetary disk. We find that, unlike in the case of a vortex at a planetary gap edge, non-axisymmetric dust trapping can be maintained for very long periods of time. We also carry out radiative transfer calculations to produce simulated sub-mm images from our results.

The outline of this paper is as follows. In Section \ref{sec:setup}, we describe the setup for our high-resolution numerical simulations, as well as the method for creating simulated mm images from them. In Section \ref{sec:results}, we present the main results of the numerical simulations, as well as the simulated images. Finally, in Section \ref{sec:discussion}, we discuss and contextualize our results.

\section{Numerical Setup}
\label{sec:setup}

We consider a two-dimensional thin disk of gas and dust described in polar coordinates $(r,\phi)$ by gas surface density $\Sigma_\mathrm{g}$, dust surface density $\Sigma_\mathrm{d}$, gas velocity $\mathbf{v}_\mathrm{g}$ and dust velocity $\mathbf{v}_\mathrm{d}$, around a star of mass $M_*$. The equation of state for the gas is locally isothermal, $P = c_\mathrm{s}^2(r)\Sigma_\mathrm{g}$, where $P$ is the height-integrated pressure and $c_\mathrm{s}(r) = c_{\mathrm{s,0}}(r/r_0)^{-1/4}$ is the radially-dependent sound speed. The scale height of the disk is $H = c_\mathrm{s}/\Omega_\mathrm{K}$, where $\Omega_\mathrm{K} = (GM_*/r^3)^{1/2}$ is the Keplerian orbital frequency. The value of $c_{\mathrm{s},0}$ is chosen so that $H/r = 0.05$ at $r_0$, and the disk is slightly flared, with $H/r \propto r^{1/4}$.

We solve the two-fluid hydrodynamic equations describing the coupled evolution of gas and dust using the \textsc{la-compass} code (Li et al.~2005, 2009; Fu et al.~2014a, 2014b),  including the effects of aerodynamic drag on the dust, as well as on the gas (i.e., the back-reaction or feedback), and dust diffusion. The radial extent of the computational domain is $[0.2, 4.39]$, in units of the scaling radius $r_0$. The initial surface density profie is $\Sigma_\mathrm{g} = \Sigma_0(r/r_0)^{-1}$, which corresponds to a steady state for a disk with constant $\alpha$ and our chosen sound speed profile. Initially, the dust surface density follows the gas surface density according to $\Sigma_\mathrm{d} = \eta_\mathrm{d}\Sigma_\mathrm{g}$. Our standard value for the initial dust-to-gas ratio is $\eta_\mathrm{d} = 0.01$. The value of this parameter, along with several others described in this section, were varied in several parameter study runs, which are defined in Table \ref{tab:parameters}. The gas surface density and velocity are fixed at their initial values at both the inner and outer boundaries. An outflow inner boundary condition and a zero radial velocity outer boundary condition are imposed on the dust. We do not consider the effects of disk self-gravity or the ``indirect potential'' due to the motion of the central star (this is justified by our small disk mass, $M_\mathrm{disk}/M_* \approx 10^{-4}$; see Section \ref{subsec:gas_dust_parameters}). Throughout this paper we express time in units of ``orbits'', where $1$ orbit $= 2\pi/\Omega_0$ is the Keplerian orbital period at $r_0$.

Our numerical resolution is $N_r \times N_\phi = 2048 \times 3072$ (with uniform grid spacing in both $r$ and $\phi$), so that the disk scale height is resolved by about $25$ grid cells at $r_0$, and each run is evolved for $10^4$ orbits. The Standard run was also simulated at a higher resolution, $N_r \times N_\phi = 4096 \times 6144$, for $4000$ orbits, for comparison.

\subsection{Viscosity Profile}

The gas has a (turbulent) kinematic viscosity given by $\nu = \alpha c_\mathrm{s}^2/\Omega_\mathrm{K}$ (Shakura \& Sunyaev 1973). The DZ is modeled using a radially varying viscosity parameter,
\be
\alpha(r) = \alpha_0\left\{1 - \frac{1}{2}\left(1 - \frac{\alpha_\mathrm{DZ}}{\alpha_0}\right) \left[1 - \tanh\left(\frac{r-r_\mathrm{DZ}}{\Delta_\mathrm{DZ}}\right)\right]\right\}.
\ee
Here $\alpha_0 = 10^{-3}$ and $\alpha_\mathrm{DZ} = 10^{-5}$ are the active zone and dead zone viscosity parameters (these values are broadly consistent with recent non-ideal MHD simulations, e.g., Simon et al.~2015, and models motivated by such simulations, e.g., Bai 2016). Note that our results are not strongly dependent on the value of $\alpha_\mathrm{DZ}$, as long as it is much less than $\alpha_0$. The transition between the two regions is located at $r_\mathrm{DZ} = 1.5 r_0$, and has a width of $\Delta_\mathrm{DZ}$, which in our standard run is chosen to be equal to the local scale height, $\Delta_\mathrm{DZ} = H(r_\mathrm{DZ}) = 0.083 r_0$.

The viscous timescale is (e.g., Lynden-Bell \& Pringle 1974) $t_\mathrm{visc} = (4/9)r^2/\nu = 2.8 \times 10^4 (r/r_0)$ orbits (for $r/r_0 \gtrsim 1.5$). Therefore, the $10^4$ orbit duration of our simulations is not long enough for significant global evolution of the gas surface density profile to occur. The viscous timescale associated with the viscosity transition, i.e., the timescale for bump accumulation, is smaller by a factor of $(\Delta_\mathrm{DZ}/r)^2$, resulting in $t_\mathrm{bump} \approx 100$ orbits.

\subsection{Gas and Dust Parameters}
\label{subsec:gas_dust_parameters}

We adopt the observationally derived parameters for the disk around Oph IRS 48, choosing $M_* = 2 M_\odot$, $r_0 = 50 \mathrm{AU}$ (so that $1$ orbit $= 250$ yr), and $\Sigma_0 = 5.401 \times 10^{-6} M_*/r_0^2 = 0.0384 \mathrm{g}/\mathrm{cm}^2$ (Bruderer et al.~2014). Our choice of $r_0$, along with our adopted DZ parameters (edge at $75 \mathrm{AU}$), result in vortices that form at about $60 - 65 \mathrm{AU}$ from the central star, coincident with the location of the peak mm emission in the system. The dust dynamics are characterized by the Stokes number, which, in the Epstein regime, is given by $\mathrm{St} = \pi s_\mathrm{p} \rho_\mathrm{p}/(2\Sigma_\mathrm{g})$, where $s_\mathrm{p}$ and $\rho_\mathrm{p}$ are the dust size and density. We choose $\rho_\mathrm{p} = 0.8 \mathrm{g}/\mathrm{cm}^3$ and $s_\mathrm{p} = 1 \mathrm{mm}$, resulting in $\mathrm{St} = 3.27$ at $r_0$ initially. Thus, the dust experiences nearly the maximum possible radial drift (the maximum occurs for $\mathrm{St} = 1$), and is also nearly the most susceptible to being trapped in vortices, hence giving the strongest possible feedback to the gas.

The total gas mass is $M_\mathrm{g} = 0.30 M_\mathrm{J}$ and the total dust mass is $M_\mathrm{d} = 0.94 M_\oplus$, although these values are sensitive to the extent of the disk, especially the location of the outer boundary, which is somewhat arbitrary. For the canonical parameters, the time required for dust to drift from $r_\mathrm{out}$ to $r = 1.25$, the approximate location where vortices form, is $t_\mathrm{drift} = 1250$ orbits. There is about $0.71 M_\oplus$ of dust beyond the vortex radius, which could potentially be trapped there. As the duration of our simulation is about $8$ times the drift timescale, we expect that the dust will have fully settled into a global quasi-equilibrium by the end of the simulation, i.e., drift on a global scale will no longer be occurring.

\subsection{Synthetic Observations}

We use the results of our hydrodynamic simulations to produce simulated maps of dust continuum emission, using \textsc{radmc-3d} (Dullemond 2012), following the method described by Jin et al.~(2016), which we briefly summarize here. First, we perform a thermal Monte Carlo simulation to determine the temperature structure of the disk, which is determined by the distribution of small ($\mu\mathrm{m}$) dust grains. For our standard parameters, $\mu\mathrm{m}$-sized dust grains have $\mathrm{St} \sim 0.003$ at $r_0$, and so do not necessarily remain perfectly coupled to the gas over thousands of orbits. Nonetheless, we assume that their distribution follows the three-dimensional structure of the gas, which is extrapolated from the gas surface density, with a fixed dust-to-gas ratio. The dust opacity for $\mu\mathrm{m}$-sized dust used in the temperature calculation is modeled as in Isella et al.~(2009), using a grain size distribution $n(a) \propto a^{-3.5}$ between $0.005 \mu\mathrm{m}$ and $100 \mu\mathrm{m}$. The disk is illuminated by a star with $T_\mathrm{eff} = 9500 \mathrm{K}$ and $L = 24 L_\odot$ (using the parameters for IRS 48 from Follette et al.~2015). Next, the two-dimensional surface density of large (mm) dust is also extrapolated into a three-dimensional density. These particles generally have a different scale height than the gas due to vertical settling. If settling is opposed only by turbulent diffusion, the dust scale height is $H_\mathrm{d} = \sqrt{\alpha/(\mathrm{St}+\alpha)} H_\mathrm{g}$ (e.g., Birnstiel et al.~2016), which can be extremely small compared to the gas scale height. However, in the scenario we consider, vertical motions associated with vortices may not allow dust to settle to this degree. Instead, we adopt $H_\mathrm{d} = 0.1 H_\mathrm{g}$, based on the results for vertical dust distributions in vortices from M{\'e}heut et al.~(2012b), for the largest particles they considered ($\mathrm{St} = 0.5$). Finally, the density distribution of mm dust, along with its opacity (calculated the same way as for the $\mu\mathrm{m}$ dust, but with a maximum grain size of $1 \mathrm{mm}$), and the temperature distribution, are used to create a synthetic thermal emission map at $440 \mu\mathrm{m}$ (ALMA Band $9$). The system is placed at a distance of $120 \mathrm{pc}$ with an inclination of $50^\circ$ (again adopting the parameters of IRS 48), and the emission map is convolved with a Gaussian beam of a given size in order to simulate an interferometric observation.

\begin{table}
\begin{center}
\begin{tabular}{|c|c|c|c|}
\hline
Name        & $\Sigma_0$ [$\mathrm{g}/\mathrm{cm}^2$] & $\eta_\mathrm{d}$ & $\Delta_\mathrm{DZ}$ \\ \hline
Standard   & $0.0384$                                                        & $0.01$                 & $H$                            \\
GasHigh   & $0.3840$                                                         & $0.01$                 & $H$                            \\
DustHigh & $0.0384$                                                         & $0.10$                 & $H$                             \\
ViscSharp & $0.0384$                                                         & $0.01$                 & $H/2$                         \\
ViscBroad & $0.0384$                                                         & $0.01$                 & $2H$                           \\ \hline
\end{tabular}
\end{center}
\caption{Simulation names and parameters}
\label{tab:parameters}
\end{table}

\section{Results}
\label{sec:results}

\begin{figure*}
\begin{center}
\includegraphics[width=0.99\textwidth,clip]{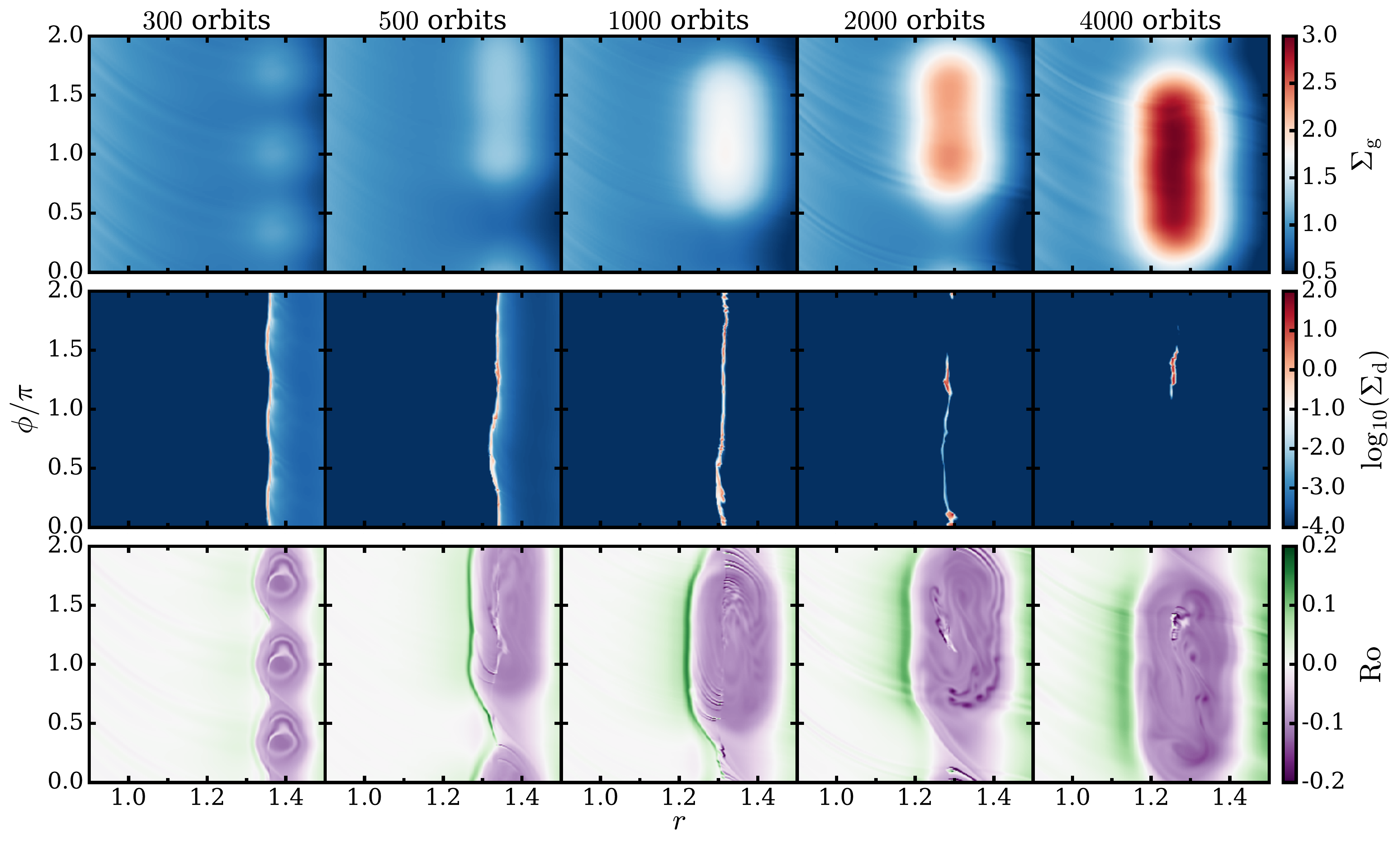}
\caption{Snapshots of gas surface density $\Sigma_\mathrm{g}$ (in units of $\Sigma_0$; top row), dust surface density $\Sigma_\mathrm{d}$ (also in units of $\Sigma_0$; middle row), and dimensionless vorticity or Rossby number, $\mathrm{Ro}$ (bottom row), in the $r-\phi$ plane, for the Standard run. Here the full azimuthal extent of the disk is shown, but only a small fraction of the radial domain, centered on the vortex region, is shown. Note that $\Sigma_\mathrm{d}$ is shown on a logarithmic scale.}
\label{fig:evolution}
\end{center}
\end{figure*}

\begin{figure}
\begin{center}
\includegraphics[width=0.49\textwidth,clip]{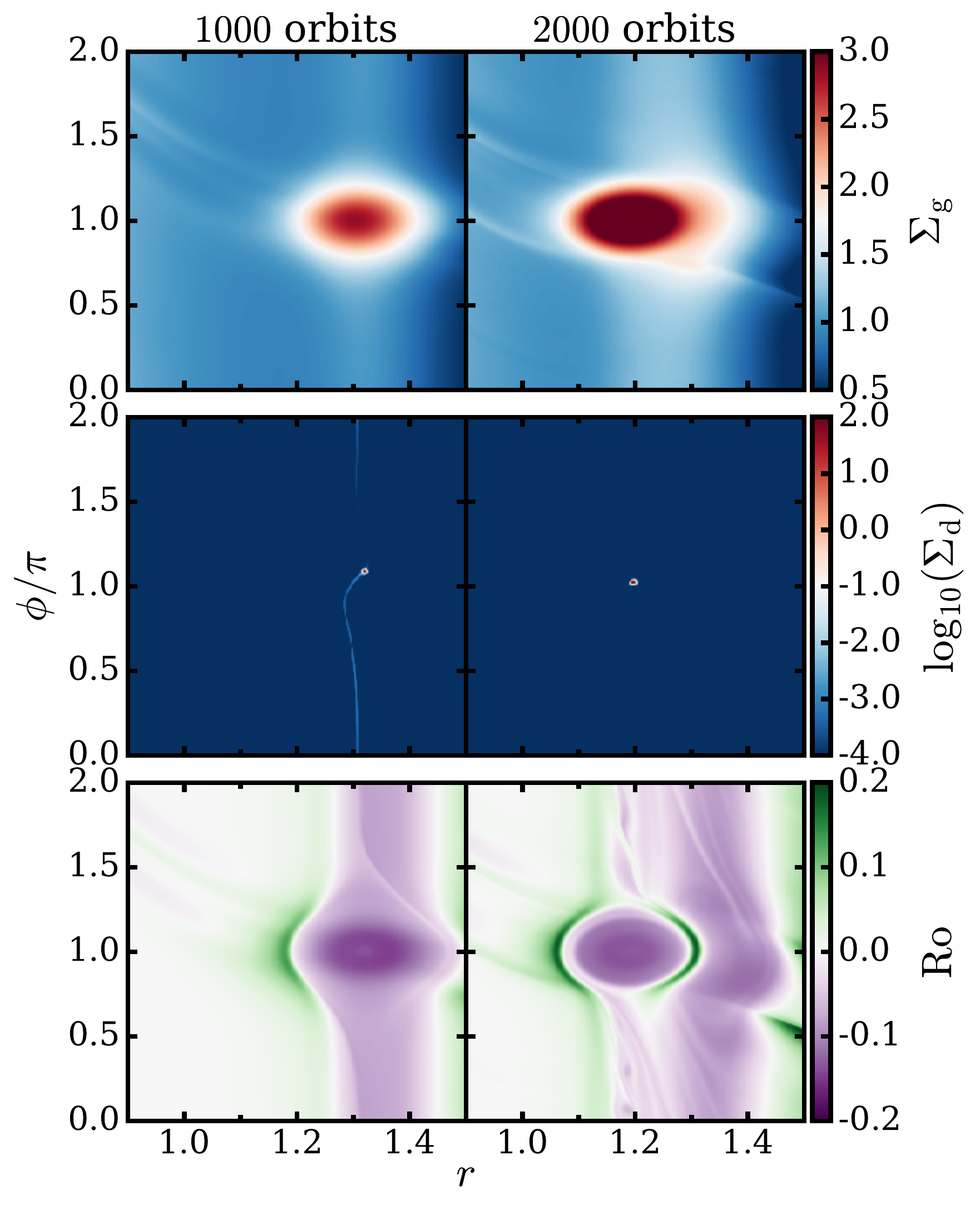}
\caption{Same as Figure \ref{fig:evolution}, with dust feedback turned off.}
\label{fig:nofeedback}
\end{center}
\end{figure}

\begin{figure}
\begin{center}
\includegraphics[width=0.49\textwidth,clip]{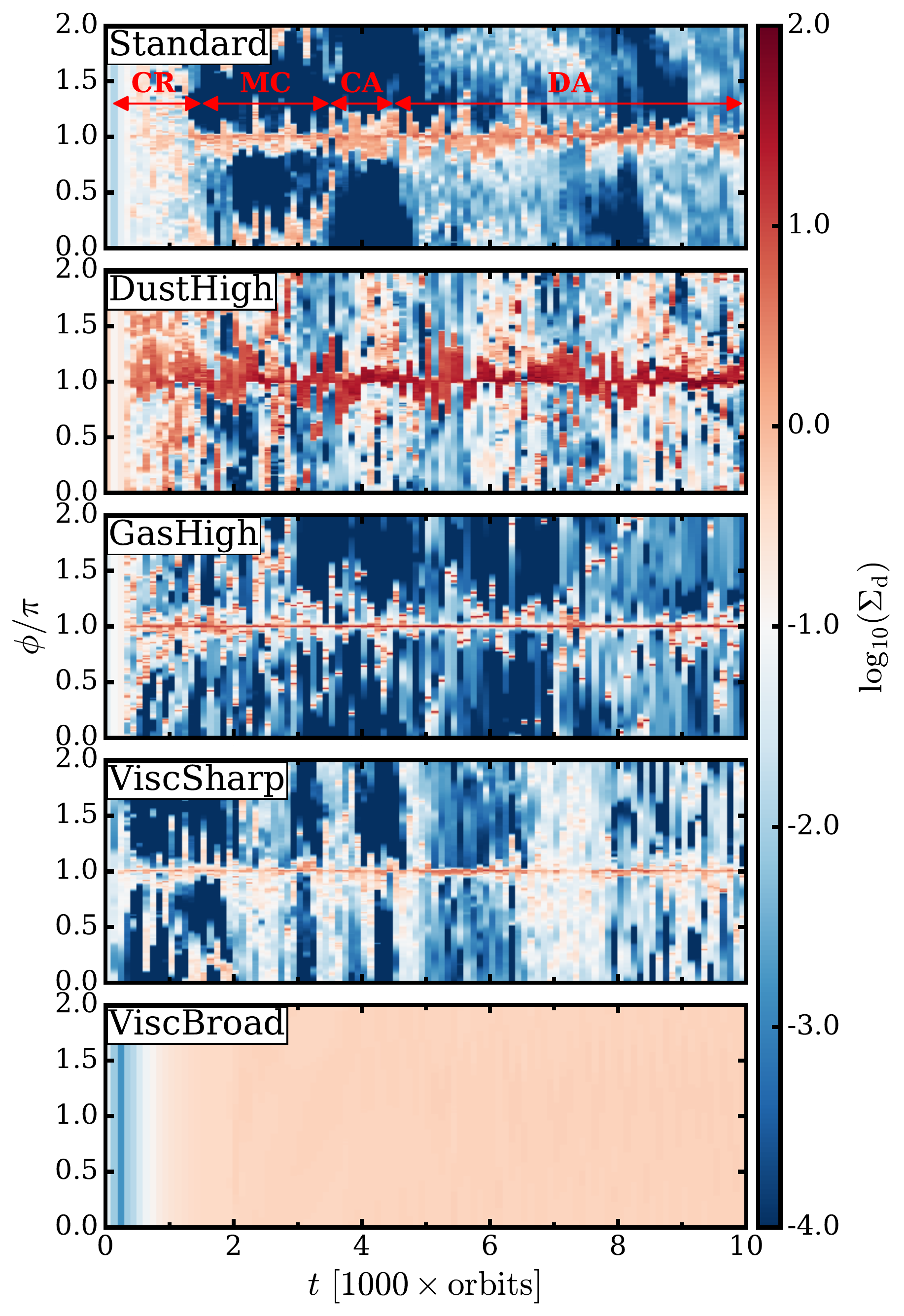}
\caption{Azimuthal dust distribution as a function of time, for different runs. In each panel, vertical slices represent the average dust density distributions in an annulus of width $H/2$, centered on the grid cell containing the largest dust density, at different times (the profiles are shifted so that the maximum conincides with $\phi = \pi$). For the Standard run, the time intervals corresponding to the clumpy ring (``CR''), multiple clumps (``MC''), clean asymmetry (``CA''), and dirty asymmetry (``DA'') phases are indicated.}
\label{fig:dust_phi_t}
\end{center}
\end{figure}

\subsection{Standard Run}

Figure \ref{fig:evolution} summarizes the evolution of the Standard run (see Table \ref{tab:parameters}). Here, snapshots of gas surface density, dust surface density, and Rossby number (dimensionless vorticity perturbation) $\mathrm{Ro} = [\mathbf{\nabla} \times (\mathbf{v}_\mathrm{g} - \mathbf{v}_\mathrm{K})]_z/(2\Omega_\mathrm{K})$ are shown at several key points in time. Note that negative values of $\mathrm{Ro}$ correspond to local rotation in the opposite direction of the bulk orbital motion, i.e., anticyclonic rotation (only anticyclonic vortices are stable against the Keplerian shear of the disk). In the first snapshot ($t = 300$ orbits), three vortices have been formed by the RWI, which has been triggered at the bump formed by accumulation of mass near the viscosity transition. A significant amount of dust has been collected axisymmetrically at the bump, such that $\Sigma_\mathrm{d}/\Sigma_\mathrm{g} \approx 1$. In the next snapshot ($500$ orbits), the three gas vortices have merged into a single vortex. The ring of dust at the pressure maximum has become clumpy, with $\Sigma_\mathrm{d}/\Sigma_\mathrm{g} \approx 10$ at some points. In the third snapshot ($1000$ orbits), the configuration is not much different than the previous one (although by this point, the outer disk has been entirely cleared of dust due to radial drift). In the fourth snapshot ($2000$ orbits), the amplitude of the gas density perturbation has increased due to continued accumulation of gas from the viscosity transition. At this point, much of the dust in the ring has become concentrated into two clumps, which are separated in azimuth. Finally, in the fifth snapshot ($4000$ orbits), the dust has become mostly concentrated into a single compact clump, in which the dust-to-gas ratio is in excess of $100$ (although the location of the largest dust density is not exactly coincident with that of the largest gas density, which is also true in some of the previous snapshots). Additionally, the peak gas density has modestly increased since the last snapshot ($\sim 20\%$ in $2000$ orbits), due to continued accretion. Note that at later times, the dust becomes somewhat more spread out again, although its distribution still remains strongly non-axisymmetric.

Figure \ref{fig:nofeedback} shows several snapshots of the Standard run with feedback turned off, in order highlight its effects. Comparing with Figure \ref{fig:evolution}, we see that feedback weakens the gas vortex, i.e., reduces the amplitude of the gas density and vorticity perturbations, and causes it to become elongated. Without feedback, the vortex aspect ratio (determined by the shape of the $\mathrm{Ro} = 0$ contour bounding the region of anticyclonic vorticity) is $\sim 8$ at $2000$ orbits, with feedback it is $\sim 20$. Since the vortex is much stronger without feedback, it migrates substantially (Li et al.~2001; Paardekooper et al.~2010), moving inward by about $5 \mathrm{AU}$ in $1000$ orbits. Thus, feedback weakens the vortex so that it migrates negligibly, staying in nearly the same place for $10^4$ orbits. The ``turbulent'' features in the vortex, traced by the vorticity field, are caused by the feedback, as without it, the vorticity distribution is much smoother. The ``shock'' feature at which the vorticity changes sign, effectively defining the edge of the vortex, is present regardless of whether or not feedback is turned on. Without feedback, the dust is very efficiently accumulated in the vortex, collecting into what is essentially a single point by $1000$ orbits, and remaining that way indefinitely. This is in stark contrast to the case with feedback, for which the dust is still in the form of a clumpy ring at $1000$ orbits, and does not accumulate into a single feature (which is more extended than in the no feedback case) until $4000$ orbits. Feedback therefore slows down and reduces the efficiency of dust trapping, but does not completely inhibit it.

The evolution of the azimuthal dust distribution in the vortex region for the Standard run (with feedback) is further illustrated in Figure \ref{fig:dust_phi_t} (top panel). We use this to define several different phases or morphologies that the dust can exhibit. Between a few hundred and about $2000$ orbits, the dust distribution is not completely axisymmetric, but rather is clumpy. We denote this the ``clumpy ring'' phase, which is represented by the first three columns of Figure \ref{fig:evolution}. Between about $2000$ and $4000$ orbits, in what we denote the ``multiple clumps'' phase, which illustrated by the fourth column of Figure \ref{fig:evolution}, there are two distinct clumps, initially separated by $\sim 180^\circ$. The two clumps eventually merge, leading to the ``clean asymmetry'' phase (e.g., the fifth column of Figure \ref{fig:evolution}), which persists for about $1000$ orbits. In this phase, essentially all of the dust in the annulus is contained in this clump. For the remainder of the evolution (after about $5000$ orbits), there is still one strong feature, but it is partially surrounded by a residual ring with a lower surface density. We denote this the ``dirty asymmetry'' phase. A snapshot of this phase, which appears to be the typical state for many thousands of orbits, is shown in Figure \ref{fig:comparison}, at $10^4$ orbits. Note that the distinction between the different morphologies is somewhat subjective and not concrete. For example, two discrete clumps of dust closely separated in azimuth tend to appear as a single feature in interferometric images (see Section \ref{subsec:images}), and so there exists a continuum between the multiple clumps and a clean asymmetry morphologies. Nonetheless, this classification scheme is useful for characterizing our results.

In addition to the asymmetric features we have focused on, there is also an axisymmetric ring of dust at about $r = 2.4$ ($120 \mathrm{AU}$). The ring forms as a result of a weak pressure maximum that forms there, in response to the pressure minimum formed just outside of the DZ edge. Once the ring is formed, dust that drifts from the outer disk gets trapped in the ring and does not make it to the vortex region. Thus, there is a finite supply of dust that can be trapped in the vortex region. A larger computational domain with more dust available in the outer disk would result in more dust being trapped in the ring, but the vortex region would be unaffected. Together, the ring beyond the DZ edge and the asymmetric features interior to it will make up the main features of our simulated images (see Section \ref{subsec:images}).

\begin{figure}
\begin{center}
\includegraphics[width=0.49\textwidth,clip]{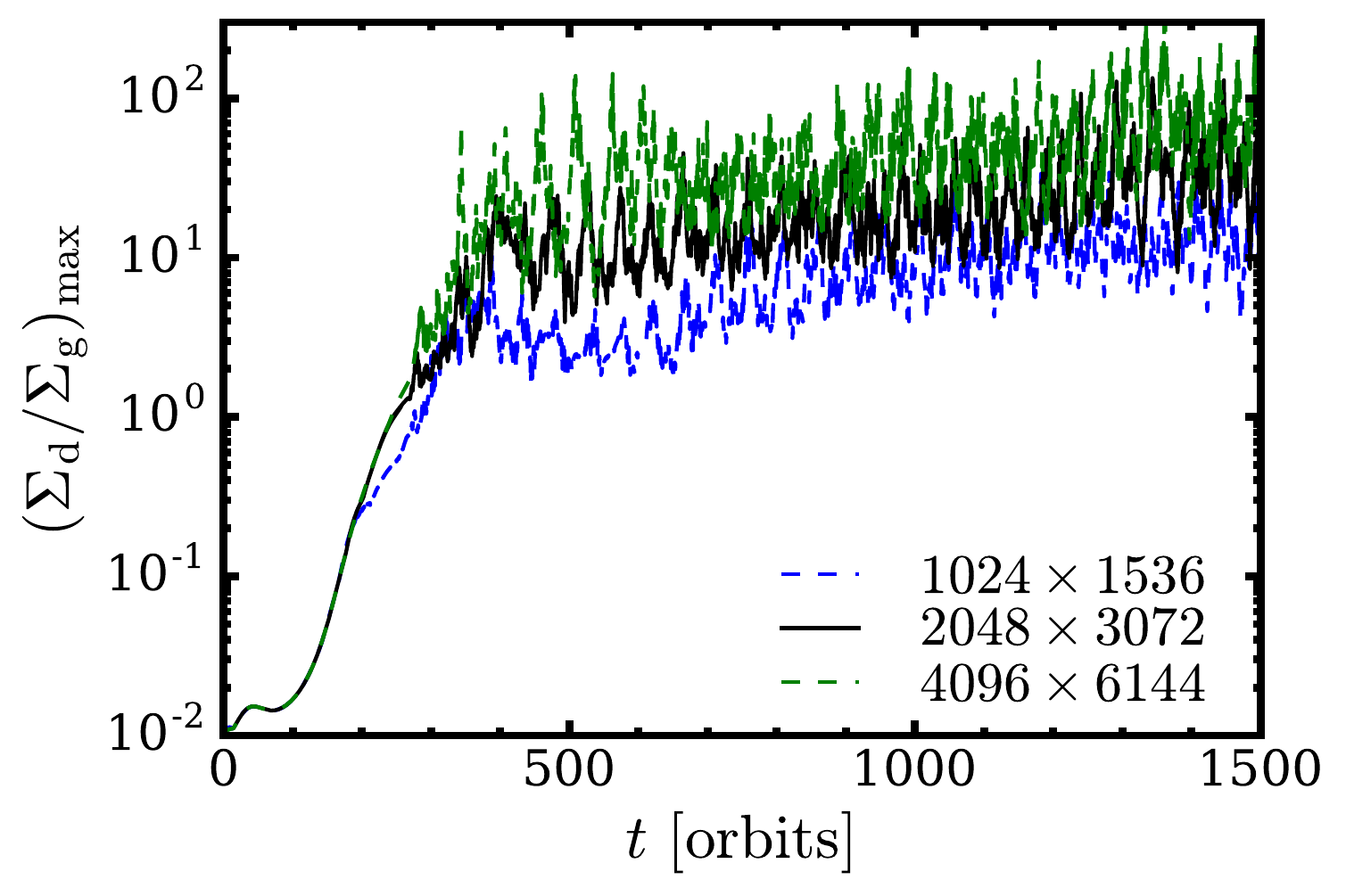}
\caption{The maximum dust-to-gas ratio versus time for the Standard run, for several numerical resolutions.}
\label{fig:resolution}
\end{center}
\end{figure}

We investigate the numerical convergence of our results in Figure \ref{fig:resolution}, which shows the maximum dust-to-gas ratio during the initial evolution of the Standard run, for several different resolutions. In all cases there is an approximately exponential rise as dust is rapidly collected in the bump/vortices, followed by fluctuations around an approximately constant level, which is always in excess of unity. Larger maximum values are achieved with increasingly higher resolution. The range of the fluctuations is similar for the two highest resolutions ($2048 \times 3072$ and $4096 \times 6144$), while for the lowest resolution ($1024 \times 1536$), the dust-to-gas ratio typically remains significantly below this range. However, even for the highest resolution, for which case the dust-to-gas ratio can reach several hundred, and thus feedback is very strong, the overall results are unchanged: the dust exhibits significant azimuthal asymmetries for the duration of the simulation.

\subsection{Parameter Dependence}

\begin{figure*}
\begin{center}
\includegraphics[width=0.99\textwidth,clip]{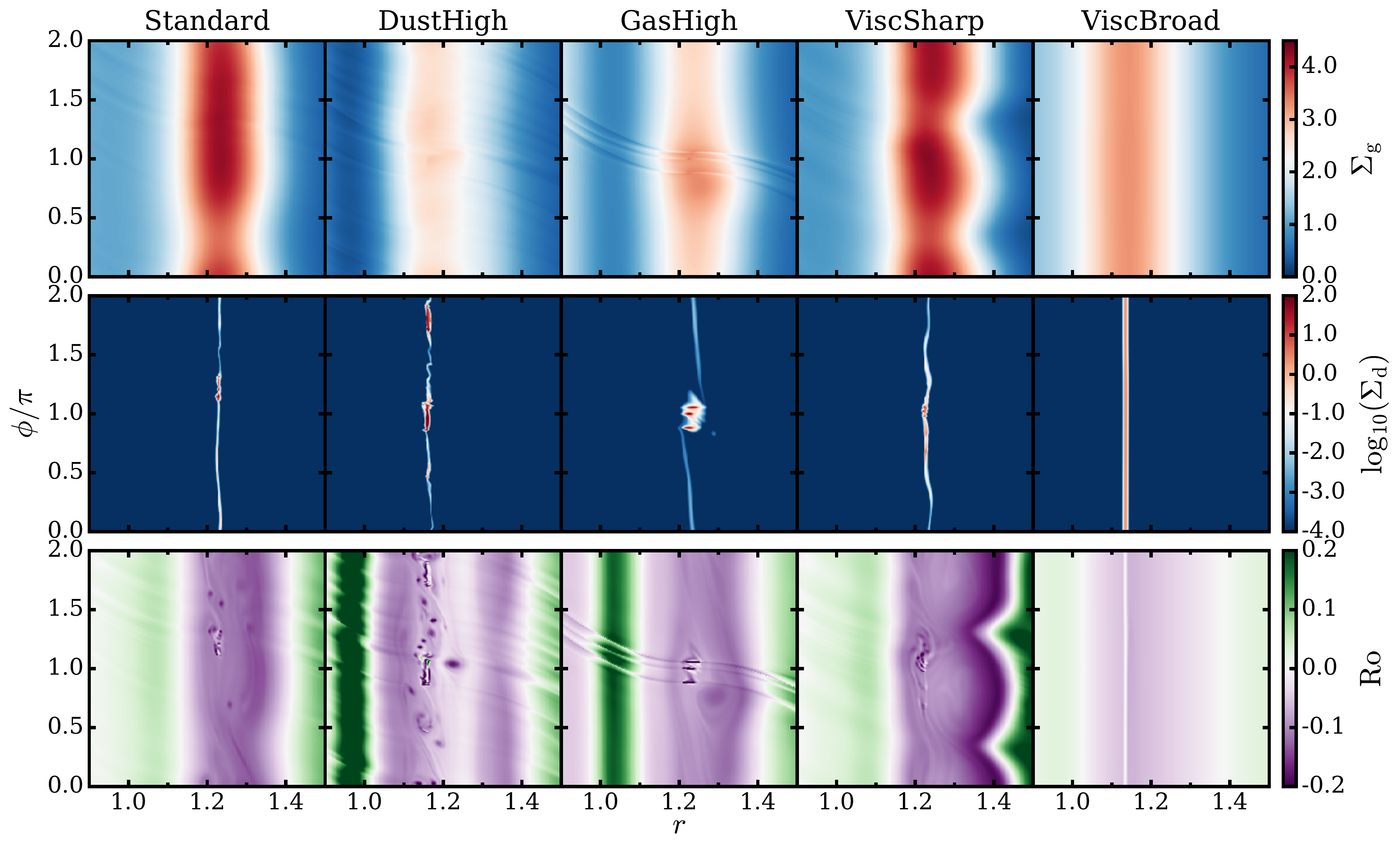}
\caption{Snapshots of different runs at $10^4$ orbits.}
\label{fig:comparison}
\end{center}
\end{figure*}

We investigated the effects changing several of the parameters of the Standard run. The parameters for these runs are given in Table \ref{tab:parameters}. Feedback is included in all of these runs. The main results are shown in Figure \ref{fig:dust_phi_t}, which illustrates the evolution of the azimuthal dust distribution in the vortex region, and Figure \ref{fig:comparison}, which shows snapshots at $10^4$ orbits.

\subsubsection{Dust-to-Gas Ratio}

A number of transition disks have been found to have enhanced dust-to-gas ratios, as large as $0.1$, compared the primordial disks, for which the typical value is $0.01$. We performed a simulation with $\eta_\mathrm{d} = 0.1$ ($10$ times larger than in the Standard run), labeled ``DustHigh''. Note that enhancement occurs in part due to radial drift of dust, while the initial condition in our simulation essentially represents a disk for which no drift has occured. Thus, there is not necessarily a direct correspondence between an observed dust-to-gas ratio and our value of $\eta_\mathrm{d}$.

As there is a larger reservoir of dust mass available to be trapped by the vortex, larger maximum dust surface densities are reached (a few times larger than in the Standard run). As a result, dust feedback is stronger, making dust concentration more difficult. This is evident in the azimuthal dust profile shown in Figure \ref{fig:dust_phi_t}: the ``clean asymmetry'' is only seen briefly, around $3500 - 4000$ and $5000 - 6000$ orbits. Instead, the distribution can usually be described by the multiple clumps or dirty asymmetry morphology. Nonetheless, the dust distribution remains asymmetric at all times.

\subsubsection{Disk Mass}

The run labeled ``GasHigh'' features a gas surface density $10$ times larger than in the canonical run (the dust surface density is also larger so that the standard value of $\eta_\mathrm{d}$ is maintained). As the standard gas surface density profiles represents IRS 48 as it is observed today, at an age of $8 \mathrm{Myr}$, after experiencing significant evolution (e.g., due to spreading, accretion, disk winds, photoevaporation, etc.), this profile could represent the system at an earlier time when the disk was more massive. The initial Stokes number of $1 \mathrm{mm}$ dust at $r_0$ is $0.327$ in this run, and so the dust dynamics are different than in the Standard run. Note that this run can also represent the dynamics of particles ten times smaller ($0.1 \mathrm{mm}$) with the standard gas density profile.

Owing to the different dust dynamics, the dust is trapped in the vortex differently, typically forming several ($2 - 4$) compact clumps, usually separated by not more than $180^\circ$ in azimuth, and often much less, as seen in Figure \ref{fig:dust_phi_t}. The clean or dirty asymmetry morphologies are rarely seen. The most striking difference compared to the Standard run is that the axisymmetric dust ring at $2.4 r_0$ is not formed. Since the dust in the outer disk has $\mathrm{St} \approx 1$, the time required for dust to drift from the outer boundary to $r \approx 1.25 r_0$ is only about $300$ orbits, about $4$ times faster than in the Standard run. This is shorter than the time required to form the weak secondary pressure maximum beyond the DZ edge, thus dust never has a chance to be trapped there, and can only be trapped by the vortex. This leads to a situation in which nearly all of the dust in the disk is situated on one side of the star at at any given time.

\subsubsection{Viscosity Transition Width}

We vary the width of the viscosity transition, in the runs labeled ``ViscSharp'' and ``ViscBroad'', which have $\Delta_\mathrm{DZ} = H/2$ and $2H$, respectively. In the ViscSharp run, the sharper viscosity gradient leads to more vigorous accumulation of mass at the DZ edge, which replenishes the RWI-unstable bump more quickly, resulting in more robust vortex formation. In the snapshot shown in Figure \ref{fig:comparison}, there are actually two vortices present, indicating that they are continuously being formed (i.e., RWI modes with $m > 1$ are still being excited). The vortex is therefore less susceptible to destruction by feedback. The amplitude of the gas perturbation is larger (although it is still not strong enough to migrate substantially), and most of the dust tends to be maintained in the center of a vortex. The evolution of the azimuthal dust distribution is not very different from the Standard run, except that the dirty asymmetry morphology is dominant most of the time.

In the ViscBroad run, mass is not accumulated into a sharp enough bump to trigger the RWI, so the disk remains axisymmetric (this result is in agreement with previous studies which find that excitation of the RWI requires $\Delta_\mathrm{DZ} \lesssim 2H$, e.g., Lyra et al.~2009; Reg{\'a}ly et al.~2012). However, at around $1800$ orbits, there is a transient growth of a very small deviation from axisymmetry in the gas that lasts for $200$ orbits and reaches an amplitude of only $\sim 1\%$, which is then damped. In response, the dust becomes slightly asymmetric (barely discernible in Figure \ref{fig:dust_phi_t}), although the asymmetry decays with time. Visually, the dust distribution near the DZ has a ring morphology for the entire simulation. Additionally, in this run, while a ring instead of an asymmetry is formed at $r \approx 1.25$, the second ring at $r \approx 2.4$, which appears in the other runs, is not created. The reason for this is similar to the reason for the lack of a ring in the GasHigh run--dust drifts through the outer disk before the secondary bump is created, in this case due to the increased timescale for gas surface density evolution, owing to the broader viscosity transition.

\subsubsection{Initial Surface Density Profile}

In all of the runs we have presented, we adopted smooth initial profiles for the gas and dust surface densities. An RWI-unstable bump is self-consistently produced by the DZ viscosity transition, as the disk attempts to reach an equilibrium in which the accretion rate, $\dot{M} \sim \nu \Sigma_\mathrm{g}$, is radially constant (this requires $\Sigma_\mathrm{g}$ to be enhanced by a factor of $\alpha_0/\alpha_\mathrm{DZ} = 100$ in the DZ). The smooth initial profile, which ignores the presence of the viscosity transition, may not necessarily represent the true initial state of the disk, and so we may ask how our results are affected if the disk has already had a chance to partially evolve toward this equilibrium. We explored this by first allowing the disk to evolve in 1D (i.e., with enforced axisymmetry) for several thousand orbits, before proceeding to evolve it in full 2D. We find that, as the gas surface density becomes enhanced in the DZ, a sharp bump near the viscosity transition always arises, since the viscous timescale associated with the viscosity transition is much shorter than the timescale to distribute mass farther inwards towards the DZ. Therefore, regardless of the specifics of the global profile, the viscosity transition guarantess the existence of an RWI-unstable bump which forms vortices. Radial drift ensures that dust becomes localized to the bump.

\subsection{Synthetic Images}
\label{subsec:images}

\begin{figure*}
\begin{center}
\includegraphics[width=0.99\textwidth,clip]{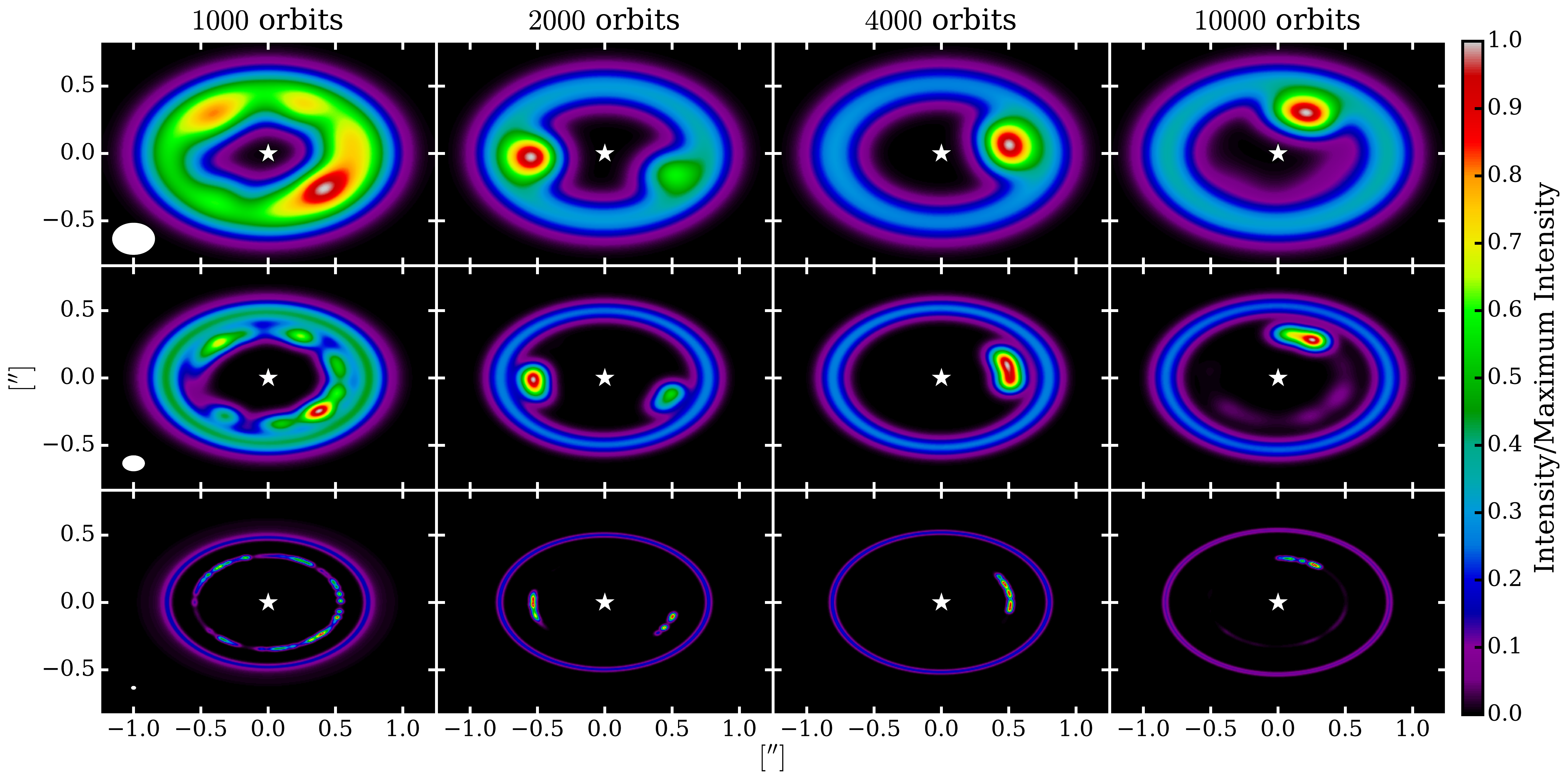}
\caption{Synthetic images at $440 \mu\mathrm{m}$ for the Standard run at several points in time. The disk is placed at a distance of $120 \mathrm{pc}$ with an inclination of $50^\circ$, and convolved with three different Gaussian beams: $0.31'' \times 0.23''$ (top row), $0.16'' \times 0.11''$ (middle row), and $0.03'' \times 0.02''$ (bottom row). The beams are shown in the leftmost panels of each row. The different snapshots correspond to several different possible morphologies, including (from left to right) the clumpy ring, multiple clumps, clean asymmetry, and dirty asymmetry phases. Note that each image has each been scaled by its maximum intensity, which ranges from about $2 - 4$ mJy/beam (for the smallest beam size) to about $15 - 20$ mJy/beam (for the largest beam size).}
\label{fig:images_evolution}
\end{center}
\end{figure*}

\begin{figure*}
\begin{center}
\includegraphics[width=0.99\textwidth,clip]{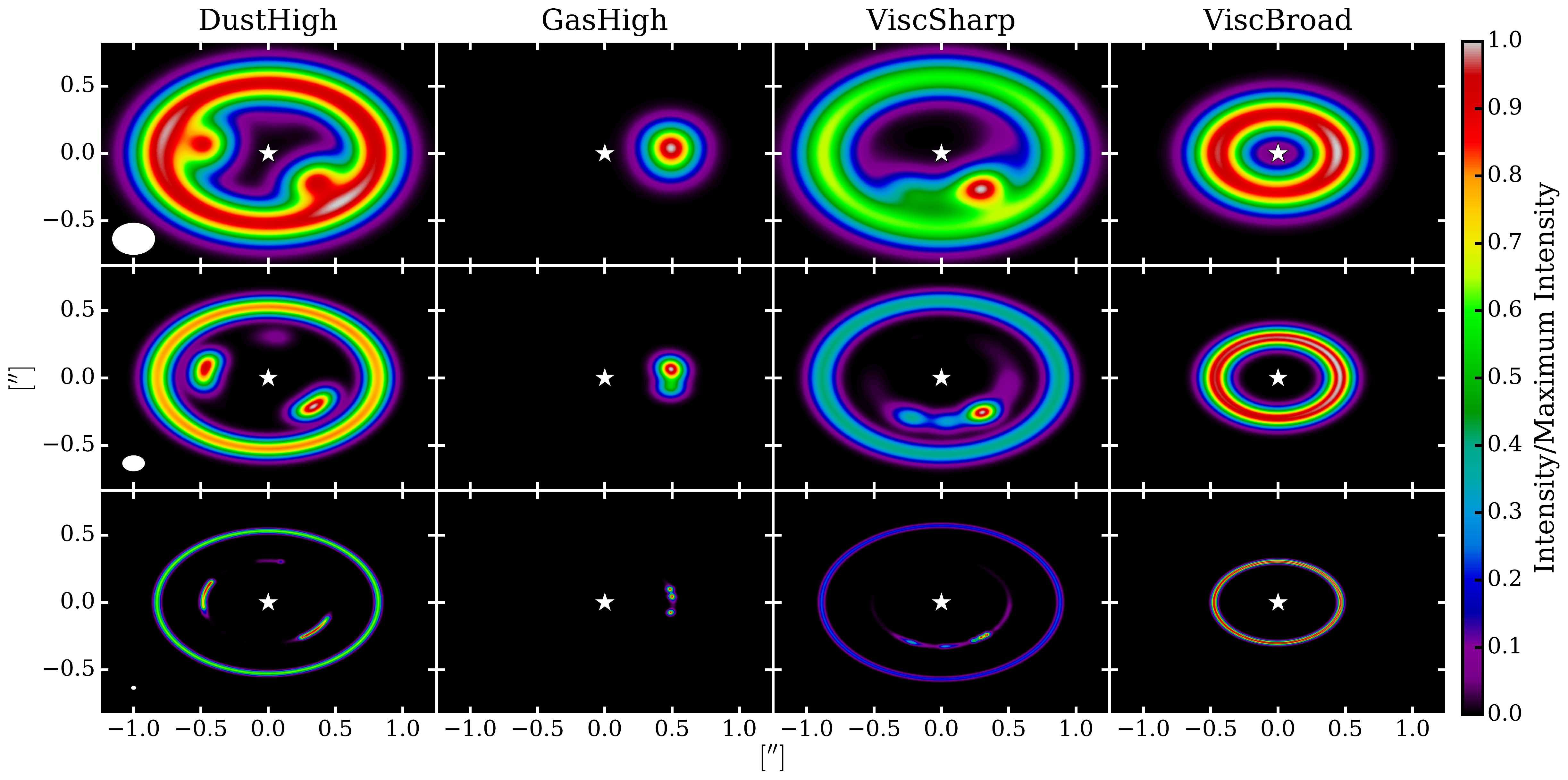}
\caption{Synthetic images at $440 \mu\mathrm{m}$ (as in Figure \ref{fig:images_evolution}) at $10^4$ orbits for the runs with varied parameters.}
\label{fig:images_comparison}
\end{center}
\end{figure*}

Images of the different phases of the Standard run are shown in Figure \ref{fig:images_evolution}, for different beam sizes. An ubiquitous feature in all of them is a ring at about $120 \mathrm{AU}$, corresponding to dust trapped at the secondary pressure maximum that arises outside of the DZ edge. The appearance of this ring does not change much during the disk evolution. Additionally, there is the asymmetric feature at about $60-65 \mathrm{AU}$, corresponding to the region shown in Figures \ref{fig:evolution} and \ref{fig:dust_phi_t}. For this run, the appearance of the dirty asymmetry morphology is not very different from that of the clean asymmetry morphology, except for an arc of emission whose brightness is about $10\%$ of the peak brightness of the asymmetric feature.

In all $4$ images, the ring and asymmetric feature are not resolvable as separate features when convolved with the largest beam, which corresponds approximately to the resolution of ALMA cycle 0 (as in the image of IRS 48 in van der Marel et al.~2013). As a result, the ring and asymmetry appear nearly coincident radially, and the brightness contrast around the apparent asymmetric ring appears to be small (a few). For a beam that is twice as small, the ring and asymmetry can be distinguished from one another, and the brightness contrast along the annulus containing the asymmetric feature is revealed to be very large--effectively infinite on the linear scale shown (in fact it is about $10^3$). At the highest resolution (corresponding approximately to the optimum resolution achievable by ALMA), it is revealed that some of the asymmetric features, which appear to constitute a single feature at lower resolutions, are in fact made of several compact clumps or arcs. For example, in the image taken at $4000$ orbits, what appeared to be a single feature at lower resolution is shown to in fact be $4$ very small patches of emission clustered together azimuthally. However, as these patches are very small (only a few grid cells wide in the hydro simulations), they are still not resolved at this resolution, and their apparent size is consistent with the beam size.

Figure \ref{fig:images_comparison} shows images of the other runs taken at $10^4$ orbits. The DustHigh run  appears similar to the Standard run at $2000$ orbits, although the ring is almost as bright as the asymmetric features, as a result of more dust being trapped there. In the GasHigh run, since the outer ring is not formed, the dust emission is dominated by the asymmetric feature, so that all of the dust appears to be on one side of the disk. For this reason, the low resolution image of this run bears the most resemblance to IRS 48. In this case, as for the Standard run, at optimum resolution, the asymmetric feature is revealed to actually consist of several unresolved sub-components. The ViscSharp image is similar to the clean asymmetry (or perhaps the multiple clumps) phase of the Standard run. The ViscBroad run exhibits only an axisymmetric ring (with $10\%$ brightness variations) at the same radius that the other runs have asymmetries. Except for the ViscBroad run, none of the runs are completely axisymmetric at $10^4$ orbits.

\section{Discussion}
\label{sec:discussion}

We performed high-resolution, two-dimensional hydrodynamic simulations of dust trapping in vortices formed at the outer edges of DZs, including the effect of dust feedback. We found that, while feedback somewhat inhibits and slows down the process of azimuthal dust trapping, it does not ultimately prevent it from occuring. This is in contrast to the case of a vortex at the edge of a planetary gap, in which the vortex is destroyed and the dust is released into a ring. The key difference between these scenarios is that in the DZ case, the viscosity transition leads to continuous accumulation of gas at the DZ edge, which allows the RWI to be sustained, constantly replenishing the vortex. This process is able to overcome dust feedback, which attemps to destroy the vortex. As a result, dust remains trapped, although the vortex is weak, and the trapping is weaker than it is without feedback. Asymmetric features in emission maps of the disk persist for at least $10^4$ orbits (the total duration of our simulations). Therefore, a disk with a DZ may appear asymmetric for most of its lifetime. We created synthetic images of thermal dust emission to compare the appearance of these features to those observed in transition disks.

Observed asymmetric features in transition disks appear to have large radial widths, much wider than the coherence width of a vortex, which is no more than a few pressure scale heights. However, the apparent widths are approximately consistent with one beam width, suggesting that the features are not resolved. Future observations may have the potential to resolve these features, which may be much narrower. This has also been proposed for vortices at planetary gap edges (Zhu \& Stone 2014), though larger features are possible due to the effects of self-gravity and the reflex motion of the central star, provided the disk is sufficiently massive (Mittal \& Chiang 2015; Zhu \& Baruteau 2016, Baruteau \& Zhu 2016). Additionally, the observed asymmetric features have different azimuthal widths, ranging from relatively compact ($\sim 45^\circ$) to extended ``horseshoe/banana'' features extending more than $180^\circ$ in azimuth. As with the radial width, the azimuthal extent may also be unresolved. Our results suggest that, if transition disk asymmetries are a result of viscosity transitions at DZ edges, they consist of very compact clumps of dust, or perhaps multiple clustered clumps, which individually are unresolvable even with a resolution of $0.02''$.

Another distinctive feature of our results is that there is also an axisymmetric ring of dust emission, located about twice as far from the star as the asymmetric features. When observed with low resolution, the two features blur together, taking the appearance of an asymmetric ring with only small brightness variations in azimuth. Higher resolution observations allow the two features to be resolved, revealing that the asymmetry in fact has a much higher brightness contrast. A similar phenomenon, in which an apparent weak asymmetry is resolved into a stronger asymmetry and an axisymmetric ring, located at different distances from the central star, was seen in observations of HD 135344B/SAO 206462. The asymmetry was first seen to be quite weak, with an azimuthal intensity variation of less than $2$ (P{\'e}rez et al.~2014), due to the (relatively) low resolution of the observation. Later observations, which achieved a better spatial resolution of $0.16''$, were able to distinguish an asymmetry with an azimuthal intensity variation of $4$ from a ring which is very close to axisymmetric, with an intensity variation of less than $1.2$ (van der Marel et al.~2016). However, the asymmetric feature is farther from the the star than the axisymmetric ring, in contrast to our results, in which the asymmetric feature is closer to the star than the ring. Additionally, this object also exhibits spiral structure in infrared scattered light, which may indicate the presence of a planet that may be responsible for the asymmetry. Nonetheless, this demonstrates the potential for future observations to distinguish a weak asymmetry from an unresolved ring/asymmetry combination.

A large amount of dust is trapped in the asymmetric features near the DZ edge. A few tenths of an Earth mass are accumulated in the the most dense clumps. As the dust densities reach about $10 \mathrm{g}/\mathrm{cm}^2$ (or several times larger if the initial dust-to-gas ratio is enhanced), gravitational collapse or the streaming instability may occur in these clumps (see Raettig et al.~2015). They contain enough mass to form not just planetesimals, but potentially planetary embryos directly. Additionally, if such embryos are created, the trapped particles, which experience maximal drift with respect to the gas, may be quickly accumulated further through pebble accretion (e.g., Owen \& Kollmeier 2016).

\section*{Acknowledgements}
Support by LANL's LDRD, UC-Fee, CSES and CNLS programs are gratefully acknowledged. All computations were carried out using LANL's Institutional Computing resources. S.~J.\@ acknowledges support from the National Natural Science Foundation of China (Grant No.~11503092) and the Strategic Priority Research Program - The Emergence of Cosmological Structures of the Chinese Academy of Sciences (Grant No.~XDB09000000).


\begin{thebibliography}{}

\bibitem[Bai \& Stone(2013)]{2013ApJ...769...76B} Bai, X.-N., \& Stone, J.~M.\ 2013, \apj, 769, 76 

\bibitem[Bai(2014)]{2014ApJ...791..137B} Bai, X.-N.\ 2014, \apj, 791, 137 

\bibitem[Bai(2016)]{2016ApJ...821...80B} Bai, X.-N.\ 2016, \apj, 821, 80 

\bibitem[Baruteau \& Zhu(2016)]{2016MNRAS.458.3927B} Baruteau, C., \& Zhu, Z.\ 2016, \mnras, 458, 3927

\bibitem[Birnstiel et al.(2016)]{2016SSRv..tmp...32B} Birnstiel, T., Fang, M., \& Johansen, A.\ 2016, \ssr,

\bibitem[Bruderer et al.(2014)]{2014A&A...562A..26B} Bruderer, S., van der Marel, N., van Dishoeck, E.~F., \& van Kempen, T.~A.\ 2014, \aap, 562, A26 

\bibitem[Casassus et al.(2013)]{2013Natur.493..191C} Casassus, S., van der Plas, G., M, S.~P., et al.\ 2013, \nat, 493, 191 

\bibitem[Cleeves et al.(2013)]{2013ApJ...772....5C} Cleeves, L.~I., Adams, F.~C., \& Bergin, E.~A.\ 2013, \apj, 772, 5 

\bibitem[Dullemond(2012)]{2012ascl.soft02015D} Dullemond, C.~P.\ 2012, Astrophysics Source Code Library, ascl:1202.015

\bibitem[Espaillat et al.(2014)]{2014prpl.conf..497E} Espaillat, C., Muzerolle, J., Najita, J., et al.\ 2014, Protostars and Planets VI, 497

\bibitem[Follette et al.(2015)]{2015ApJ...798..132F} Follette, K.~B., Grady, C.~A., Swearingen, J.~R., et al.\ 2015, \apj, 798, 132 

\bibitem[Fu et al.(2014)]{2014ApJ...788L..41F} Fu, W., Li, H., Lubow, S., \& Li, S.\ 2014, \apjl, 788, L41 

\bibitem[Fu et al.(2014)]{2014ApJ...795L..39F} Fu, W., Li, H., Lubow, S., Li, S., \& Liang, E.\ 2014, \apjl, 795, L39 v

\bibitem[Gammie(1996)]{1996ApJ...457..355G} Gammie, C.~F.\ 1996, \apj, 457, 355 

\bibitem[Isella et al.(2009)]{2009ApJ...701..260I} Isella, A., Carpenter, J.~M., \& Sargent, A.~I.\ 2009, \apj, 701, 260

\bibitem[Isella et al.(2013)]{2013ApJ...775...30I} Isella, A., P{\'e}rez, L.~M., Carpenter, J.~M., et al.\ 2013, \apj, 775, 30 

\bibitem[Jin et al.(2016)]{2016ApJ...818...76J} Jin, S., Li, S., Isella, A., Li, H., \& Ji, J.\ 2016, \apj, 818, 76 

\bibitem[Lesur et al.(2014)]{2014A&A...566A..56L} Lesur, G., Kunz, M.~W., \& Fromang, S.\ 2014, \aap, 566, A56 

\bibitem[Li et al.(2000)]{2000ApJ...533.1023L} Li, H., Finn, J.~M., Lovelace, R.~V.~E., \& Colgate, S.~A.\ 2000, \apj, 533, 1023 

\bibitem[Li et al.(2001)]{2001ApJ...551..874L} Li, H., Colgate, S.~A., Wendroff, B., \& Liska, R.\ 2001, \apj, 551, 874 

\bibitem[Li et al.(2005)]{2005ApJ...624.1003L} Li, H., Li, S., Koller, J., et al.\ 2005, \apj, 624, 1003 

\bibitem[Li et al.(2009)]{2009ApJ...690L..52L} Li, H., Lubow, S.~H., Li, S., \& Lin, D.~N.~C.\ 2009, \apjl, 690, L52 

\bibitem[Lovelace et al.(1999)]{1999ApJ...513..805L} Lovelace, R.~V.~E., Li, H., Colgate, S.~A., \& Nelson, A.~F.\ 1999, \apj, 513, 805

\bibitem[Lynden-Bell \& Pringle(1974)]{1974MNRAS.168..603L} Lynden-Bell, D., \& Pringle, J.~E.\ 1974, \mnras, 168, 603 

\bibitem[Lyra et al.(2009)]{2009A&A...497..869L} Lyra, W., Johansen, A., Zsom, A., Klahr, H., \& Piskunov, N.\ 2009, \aap, 497, 869 

\bibitem[Lyra \& Mac Low(2012)]{2012ApJ...756...62L} Lyra, W., \& Mac Low, M.-M.\ 2012, \apj, 756, 62 

\bibitem[Lyra \& Lin(2013)]{2013ApJ...775...17L} Lyra, W., \& Lin, M.-K.\ 2013, \apj, 775, 17 

\bibitem[Lyra et al.(2015)]{2015A&A...574A..10L} Lyra, W., Turner, N.~J., \& McNally, C.~P.\ 2015, \aap, 574, A10

\bibitem[Meheut et al.(2012a)]{2012A&A...542A...9M} M{\'e}heut, H., Keppens, R., Casse, F., \& Benz, W.\ 2012, \aap, 542, A9 

\bibitem[Meheut et al.(2012b)]{2012A&A...545A.134M} M{\'e}heut, H., Meliani, Z., Varniere, P., \& Benz, W.\ 2012, \aap, 545, A134 

\bibitem[Miranda et al.(2016)]{2016MNRAS.457.1944M} Miranda, R., Lai, D., \& M{\'e}heut, H.\ 2016, \mnras, 457, 1944 

\bibitem[Mittal \& Chiang(2015)]{2015ApJ...798L..25M} Mittal, T., \& Chiang, E.\ 2015, \apjl, 798, L25 

\bibitem[Owen \& Kollmeier(2016)]{2016arXiv160708250O} Owen, J.~E., \& Kollmeier, J.~A.\ 2016, arXiv:1607.08250 

\bibitem[Paardekooper et al.(2010)]{2010ApJ...725..146P} Paardekooper, S.-J., Lesur, G., \& Papaloizou, J.~C.~B.\ 2010, \apj, 725, 146

\bibitem[P{\'e}rez et al.(2014)]{2014ApJ...783L..13P} P{\'e}rez, L.~M., Isella, A., Carpenter, J.~M., \& Chandler, C.~J.\ 2014, \apjl, 783, L13 

\bibitem[Raettig et al.(2015)]{2015ApJ...804...35R} Raettig, N., Klahr, H., \& Lyra, W.\ 2015, \apj, 804, 35 

\bibitem[Ragusa et al.(2017)]{2017MNRAS.464.1449R} Ragusa, E., Dipierro, G., Lodato, G., Laibe, G., \& Price, D.~J.\ 2017, \mnras, 464, 1449  

\bibitem[Reg{\'a}ly et al.(2012)]{2012MNRAS.419.1701R} Reg{\'a}ly, Z., Juh{\'a}sz, A., S{\'a}ndor, Z., \& Dullemond, C.~P.\ 2012, \mnras, 419, 1701 

\bibitem[Shakura \& Sunyaev(1973)]{1973A&A....24..337S} Shakura, N.~I., \& Sunyaev, R.~A.\ 1973, \aap, 24, 337

\bibitem[Simon et al.(2015)]{2015MNRAS.454.1117S} Simon, J.~B., Lesur, G., Kunz, M.~W., \& Armitage, P.~J.\ 2015, \mnras, 454, 1117 

\bibitem[Surville et al.(2016)]{2016ApJ...831...82S} Surville, C., Mayer, L., \& Lin, D.~N.~C.\ 2016, \apj, 831, 82 

\bibitem[van der Marel et al.(2013)]{2013Sci...340.1199V} van der Marel, N., van Dishoeck, E.~F., Bruderer, S., et al.\ 2013, Science, 340, 1199

\bibitem[van der Marel et al.(2016)]{2016arXiv160705775V} van der Marel, N., Cazzoletti, P., Pinilla, P., \& Garufi, A.\ 2016, arXiv:1607.05775 

\bibitem[Varni{\`e}re \& Tagger(2006)]{2006A&A...446L..13V} Varni{\`e}re, P., \& Tagger, M.\ 2006, \aap, 446, L13 

\bibitem[Zhu \& Baruteau(2016)]{2016MNRAS.458.3918Z} Zhu, Z., \& Baruteau, C.\ 2016, \mnras, 458, 3918 

\bibitem[Zhu \& Stone(2014)]{2014ApJ...795...53Z} Zhu, Z., \& Stone, J.~M.\ 2014, \apj, 795, 53 

\end{thebibliography}
\end{document}